\documentclass[amsmath,preprintnumbers]{revtex4}
\usepackage{graphicx}
\usepackage{bm}
\newcommand{\pd}[2]{\frac{\partial {#1}}{\partial {#2}}} 

\newcommand{\al}{{\alpha}}

\newcommand{\ga}{{\gamma}}

\newcommand{\de}{{\delta}}
\newcommand{\De}{{\Delta}}

\newcommand{\ka}{{\kappa}}

\newcommand{\om}{{\omega}}
\newcommand{\Om}{{\Omega}}

\begin{document}
\title{New Consequences of Induced Transparency in a Double-$\Lambda$ scheme:\\
Destructive Interference In Four-wave Mixing}
\author{M.G. Payne}
\affiliation{Department of Physics, Georgia Southern University, Statesboro, GA}
\author{Lu Deng}
\affiliation{Electron and Optical Physics Division, NIST, Gaithersburg, MD}
\date{\today}
\begin{abstract}
We investigate a four-state system interacting with long and short laser
pulses in a weak probe beam approximation.  We show that when all lasers
are tuned to the exact unperturbed resonances, part of the four-wave mixing
(FWM) field is strongly absorbed.  The part which is not absorbed has
the exact intensity required to destructively
interfere with the excitation pathway involved in producing the FWM state.
We show that with this three-photon destructive interference, the conversion
efficiency can still be as high as 25\%.  Contrary to common belief,
our calculation shows that this process, where an ideal one-photon electromagnetically
induced transparency is established, is not most suitable for high efficiency
conversion.  With appropriate phase-matching and propagation distance, and
when the three-photon destructive interference does not occur, we show that
the photon flux conversion efficiency is independent of probe intensity and
can be close to 100\%.  In addition, we show clearly that the conversion efficiency is not
determined by the maximum atomic coherence between two lower excited states, as
commonly believed.  It is the combination of phase-matching and constructive
interference involving the two terms arising in producing the mixing wave that is the key
element for the optimized FWM generation. Indeed, in this scheme no appreciable excited
state is produced, so that the atomic coherence between states $|0>$\ and $|2>$\
is always very small.
\end{abstract}
\maketitle
\section{Introduction}
Efficient four-wave mixing (FWM) processes in the context of electromagnetically
induced transparency (EIT) [1] involving a double-$\Lambda$ scheme have been the
subject of several recent studies [2-4].  In almost all of these studies, a large atomic coherence has been
considered to be a key player in yielding a high conversion efficiency.
Therefore, maximum atomic coherence was
assumed.  Furthermore, all studies available so far are based on either a
steady-state approximation[2-4] or a full-numerical calculation[3]
to obtain predictions on the system. A
numerical calculation usually fails to yield as much insight into a problem
as would be obtained from an analytical solution. The steady-state treatment
is analytical, but it oversimplifies the problem and, in some cases,
it leads to incorrect conclusions.  In particular,
it is known that a steady-state treatment of a four-level double-$\Lambda$
scheme can lead to an inconsistent solution in predicting quantum destructive
interference effects that have profound relations to the wave mixing process.
In this study, we describe an  approximate analytical solution to a four-state
double-$\Lambda$ scheme that is different from those studied previously [2-4].
Three features distinguish the present study from previous works:
(1). A self-consistent fully time-dependent treatment leads to a three-photon destructive interference
that extinguishes the laser excitation to the terminal state of a three-photon
resonance, yet still provides nearly 25\% frequency conversion efficiency.
Such a three-photon destructive interference together with a high conversion
efficiency from a quenched FWM state have not previously been reported in any double-$\Lambda$
scheme in the context of EIT, and cannot be obtained from a simple steady-state
treatment. (2). We show that resonance excitation (which leads to perfect one-photon
EIT) is not the desired condition for optimum production of the mixing wave.  We
point out that sizable detunings from these resonances are required in order
to avoid the three-photon destructive interference that limits the conversion
efficiency, and (3) We show that, contrary to common beliefs, high conversion
efficiency is not determined by having the maximum atomic coherence between
the states $|0>$\ and $|1>$.  Indeed, in this four-wave mixing scheme
very little depletion of the ground state occurs and $\rho_{01}$\ remains
very small throughout the probe pulse. The correct criteria for achieving high
conversion efficiency is a combination of the conventional phase-matching
condition and a condition that ensures the constructive interference between
two terms in the expression for the FWM field. In order to make both of the
two terms of comparable size, a relatively large detuning is required to
reduce near resonance absorption of one part of the four-wave mixing field.

\section{Theoretical Model}
Consider a four-level system that interacts with two transform-limited
lasers ($E_{L1}(\om_{L1})$\ and $E_{L2}(\om_{L2})$, pulse length $\tau_0$, Fig.1).
A short pulse probe laser (pulse length $\tau<<\tau_0$) is tuned near
the $|0>\rightarrow |2>$ resonance and fired at a predetermined delay time.
We assume that during the pulse length of the probe laser the two long pulsed
lasers
at $\om_{L1}$\ and $\om_{L2}$ are sufficiently powerful to strongly saturate
$|2>\rightarrow |1>$ and $|1>\rightarrow |3>$ transitions, therefore, $|\Om_{12}\tau|>>1$ and
$|\Om_{13}\tau|>>1$.  As usual, $\Om_{ij}=D_{ij}E_{L}/(2\hbar)$ and $E_{L}$ are
the one half Rabi frequency and the amplitude of the field for the respective
transitions.  Our objective is to seek a perturbative treatment for the response
of the system to the short pulsed laser and to investigate the dynamics of the
generated wave.
\vskip 10pt
We start with three equations of motion for the amplitudes of atomic wave
function in the non-depleted ground state approximation.  We thus require that
$|\Omega_{02}|<<|\Omega_{21}|$, so that most of the population remains in
state $|0>$.  Taking $A_0\simeq{1}$, we have
\begin{subequations}
\begin{eqnarray}
\pd{A_1}{t}&=&i\Omega_{12}A_2+i\Omega_{13}A_3+i(\de_1+i\frac{\ga_1}{2})A_1, \\
\pd{A_2}{t}&=&i(\de_2+i\ga_2/2)A_2+i\Omega_{20}+i\Omega_{21}A_1, \\
\pd{A_3}{t}&=&i(\de_3+i\ga_3/2)A_3+i\Omega_{31}A_1+i\Omega_{30}.
\end{eqnarray}
\end{subequations}
In Eqs.(1), $\de_1$\ is the detuning from
the two-photon resonance between $|0>$\ and $|1>$,
$\de_2$\ is the detuning of the short pulse probe laser from
the $|0>\rightarrow |2>$\ resonance, $\de_3$ is the detuning from the
three-photon resonance involving the transition $|0>\rightarrow |3>$.  We will eventually assume
that the lasers are tuned to the two-photon resonance $|0>\rightarrow |1>$\
and that the lifetime of this state is very long. Bear
in mind that the generation of the four-wave mixing field will be shown to be very efficient,
therefore, we must solve the above three equations of motion simultaneously
with Maxwell's equations for both the probe and the generated fields.
Taking unfocused beams and introducing $\ka_{ij}=2\pi\om_{ji}N|D_{ij}|^2/(\hbar c)$
where $N$\ is the concentration in cm$^{-3}$, Maxwell's equations for these
fields in slowly varying amplitude and phase approximation can be expressed as
\begin{subequations}
\begin{eqnarray}
\pd{\Omega_{20}}{z}+\frac{1}{c}\pd{\Omega_{20}}{t}&=&i\ka_{02}A_2, \\
\pd{\Omega_{30}}{z}+\frac{1}{c}\pd{\Omega_{30}}{t}&=&i\ka_{03}A_3.
\end{eqnarray}
\end{subequations}
Notice that in the present model, there exist two very different time scales,
i.e. $\tau<<\tau_0$, therefore,
if the probe pulse occurs at the peak of the long pulse lasers, the amplitudes
of the latter will remain nearly constant through the entire probe pulse.
We therefore will be seeking a fully time dependent response of the system during the
period when the probe pulse is present while treating both long pulse laser
fields as time independent quantities.  With this method in mind, Eqs.(1) and (2)
can be solved analytically.  Taking Fourier
transform on the both sides of Eqs.(1) and (2), we obtain
\begin{subequations}
\begin{eqnarray}
\Om_{21}\al_1+(\de_2+\om+i\ga_2/2)\al_2&=&-W_{20},\\
(\de_1+\om+i\ga_1/2)\al_1+\Om_{12}\al_2+\Om_{13}\al_3&=&0, \\
\Om_{31}\al_1+(\de_3+\om+i\ga_3/2)\al_3&=&-W_{30},\\
\pd{W_{20}}{z}-i\frac{\om}{c}W_{20}&=& i\ka_{02}\al_2,\\
\pd{W_{30}}{z}-i\frac{\om}{c}W_{30}&=& i\ka_{03}\al_3.
\end{eqnarray}
\end{subequations}
where $\al_1$, $\al_2$, $\al_3$, $W_{20}$\ and $W_{30}$\ are the Fourier
transforms of $A_1$, $A_2$, $A_3$, $\Omega_{20}$\ and $\Omega_{30}$,
respectively.  Eqs. (3a-3c) can be solved in terms of
$W_{20}$\ and $W_{30}$\ with the result
\begin{subequations}
\begin{eqnarray}
\al_1&=&-\frac{D_3\Om_{12}}{\Delta}W_{20}-\frac{D_2\Om_{13}}{\Delta}W_{30},\\
\al_2&=&-\frac{\Om_{21}\Om_{13}}{\Delta}W_{30}+\frac{|\Om_{13}|^2-D_1D_3}{\Delta}W_{20},\\
\al_3&=&-\frac{\Om_{31}\Om_{12}}{\Delta}W_{20}+\frac{|\Om_{12}|^2-D_1D_2}{\Delta}W_{30},
\end{eqnarray}
\end{subequations}
where,
\begin{eqnarray}
D_1&=&\de_1+\om+i\ga_1/2,\nonumber\\
D_2&=&\de_2+\om+i\ga_2/2,\nonumber\\
D_3&=&\de_3+\om+i\ga_3/2,\nonumber\\
\Delta&=&D_1D_2D_3-D_3|\Om_{12}|^2-D_2|\Om_{13}|^2.\nonumber
\end{eqnarray}
When these equations are used in  Eqs (3d-3e), we obtain
\begin{subequations}
\begin{eqnarray}
\pd{W_{20}}{z}-i\frac{\om}{c}W_{20}&=&i\ka_{02}\frac{(|\Om_{13}|^2-D_1D_3)}{\Delta}W_{20}-i\ka_{02}\frac{\Om_{21}\Om_{31}}{\Delta}W_{30},\\
\pd{W_{30}}{z}-i\frac{\om}{c}W_{30}&=&i\ka_{03}\frac{(|\Om_{12}|^2-D_1D_2)}{\Delta}W_{30}-i\ka_{03}\frac{\Om_{31}\Om_{12}}{\Delta}W_{20}.
\end{eqnarray}
\end{subequations}

For given $W_{20}(0,\om)$ and with $W_{30}(0,\om)=0$, Eqs (5) can be solved
analytically, yielding
\begin{subequations}
\begin{eqnarray}
W_{30}(z,\om)&=&i\frac{W_{20}(0,\om)S_3}{\Lambda}e^{iDz}\sin(\Lambda z),\\
W_{20}(z,\om)&=&\frac{W_{20}(0,\om)}{\Lambda}e^{iDz}\left(i\frac{K_2-K_3}{2}\sin(\Lambda z)+\Lambda\cos(\Lambda z)\right),
\end{eqnarray}
\end{subequations}
where we have defined the new parameters
\begin{subequations}
\begin{eqnarray}
\Lambda&=&\sqrt{\left(\frac{K_2-K_3}{2}\right)^2+S_2S_3}, \\
D&=&\frac{K_2+K_3}{2},\\
K_2&=&\frac{\om}{c}+\ka_{02}\frac{|\Om_{13}|^2-D_1D_3}{\Delta}, K_3=\frac{\om}{c}+\ka_{03}\frac{|\Om_{12}|^2-D_1D_2}{\Delta},\\
S_2&=&-\ka_{02}\frac{\Om_{21}\Om_{13}}{\Delta}, S_3=-\ka_{03}\frac{\Om_{31}\Om_{12}}{\Delta}.
\end{eqnarray}
\end{subequations}

If we use, for the probe laser at the entrance to the medium,
\begin{equation}
\Om_{02}(0,t)=\Om_{02}(0,0)e^{-(t/\tau)^2},\nonumber
\end{equation}
we find
\begin{equation}
W_{02}(0,\eta)/\Om_{02}(0,0)=\frac{\tau}{\sqrt{2}}e^{-\eta^2/4},\nonumber
\end{equation}
where we have introduced the dimensionless variable $\eta=\om\tau$.  We then get
\begin{subequations}
\begin{eqnarray}
\Om_{30}\left(z,\frac{t}{\tau}\right)&=&\frac{i\Om_{20}(0,0)}{\sqrt{4\pi}}\int_{-\infty}^{\infty} d\eta e^{-\frac{\eta^2}{4}}e^{iD(\eta)z}e^{-i\eta\frac{t}{\tau}}S_3(\eta)\frac{\sin(\Lambda z)}{\Lambda},\\
\Om_{20}\left(z,\frac{t}{\tau}\right)&=&\frac{\Om_{20}(0,0)}{\sqrt{4\pi}}\int_{-\infty}^{\infty} d\eta e^{-\frac{\eta^2}{4}}e^{iD(\eta)z}e^{-i\eta\frac{t}{\tau}}\left[i\frac{K_2-K_3}{2\Lambda}\sin(\Lambda z)+\cos(\Lambda z)\right].
\end{eqnarray}
\end{subequations}

In the following section, we will discuss the physical implications of Eq.(8) by examining
some limiting cases where Eq.(8) can be evaluated analytically.
In Section IV, we will evaluates Eqs. (8) numerically and compare the
results with the analytical approximations.
\vskip 10pt
\leftline{\bf{III. Discussions}}

In this section we focus on the physical interpretation of Eqs.(8).  We will
consider some limiting cases where the inverse transform of the generated
field can be carried out analytically.  These limiting cases provide a great
deal of insight into the wave propagation effects.
\vskip 10pt
Consider the limiting case where $|\Om_{12}|, |\Om_{13}|>>|\de_1|, |\de_2|, |\de_3|,
\ga_1, \ga_2, \ga_3$. In fact, in what follows we shall always use $\de_1=0$\ and
assume that state $|1>$\ is a second hyperfine level of the ground state, so that
in an ultra-cold vapor $\ga_1\tau<<1$.  In this case the width of state $|1>$
is determined by the very slow rate of
collisions between the atoms of cold low density vapor. Under these
conditions, an accurate approximation to the parameters in Eqs. (7) can be
obtained by assuming
$|\De_1|<< |\De_2|,\quad|\De_3|$, and $|\Om_{12}|,|\Om_{13}|>>|\De_2|,|\De_3|$.
Within this limit we can expand $\Lambda$\ by making use of the assumption
that either $|\Om_{13}|^2$, or $|\Om_{12}|^2$\ is much larger than either
$|D_1D_2|$\ or $|D_1D_3|$, therefore,
\begin{equation}
\frac{K_2+K_3}{2}-\sqrt{\left(\frac{K_2-K_3}{2}\right)^2+S_2S_3}\simeq -D_1\frac{\ka_{12}\ka_{32}}{\ka_{12}|\Om_{13}|^2+\ka_{32}|\Om_{12}|^2}.
\end{equation}
If we replace $sine$\ and $cosine$\ in the inverse transform by complex
exponentials, this approximation allows us to evaluate part of the integrals
analytically for arbitrary pulse shape.  When the detunings $\de_2$\ and $\de_3$\
are very small compared with the half-Rabi frequencies, the other integrals
are damped out in a small distance due to absorption.
\vskip 10pt
\noindent We first analyze the case where $\de_1=\de_2=\de_3=0$, i.e. all lasers
are tuned exactly on the unperturbed resonances. This is the limit where
one-photon EIT is achieved.  In this limit we find that
after a large propagation distance, and at a point where the FWM has built up
sufficiently
\begin{subequations}
\begin{eqnarray}
\Om_{30}(z,t)&=&\frac{\ka_{03}\Om_{31}\Om_{12}}{\ka_{02}|\Om_{13}|^2+\ka_{03}|\Om_{12}|^2}\Om_{20}(0,t-\frac{z}{V_{g1}}), \\
\Om_{20}(z,t)&=&\frac{\ka_{03}|\Om_{12}|^2}{\ka_{02}|\Om_{13}|^2+\ka_{03}|\Om_{12}|^2}\Om_{20}(0,t-\frac{z}{V_{g1}}), 
\end{eqnarray}
\end{subequations}
where
\begin{equation}
\frac{1}{V_{g1}}=\frac{1}{c}+\frac{\ka_{12}\ka_{32}}{\ka_{12}|\Om_{13}|^2+\ka_{32}|\Om_{12}|^2}.\nonumber
\end{equation}
\vskip 10pt
This result indicates that in this limit $\Om_{20}(z,t)=(\Om_{21}/\Om_{31})\Om_{30}(z,t)$.[5]
Indeed, Eq. (4c) shows that in this limit $\al_3\simeq 0$\ for all $\om$,
providing $z$\ is large enough to make
$|\exp(i[\ka_{02}|\Om_{13}|^2+\ka_{03}|\Om_{12}|^2]/\Delta])|<<1$.
It is required that $|\Om_{12}|^2>>|D_1D_2|$\ in order for $W_{02}/W_{03}=\Om_{12}/\Om_{13}$\
to imply that $\al_3=0$. In other words, $a$ $destructive$ $interference$ $has$
$occured$ $between$ $the$ $excitation$ $pathways$ $for$ $the$ $state$. [6]  We
emphasize that this suppression of the excitation of the state $|3>$ cannot be obtained
from the usual steady-state treatment of the atomic equations of motion.  Indeed,
the condition derived in the steady-state frame work for a three-photon
destructive interference leads to an inconsistent prediction of the effect.
We have evaluated the special case where $\ka_{12}\tau=\ka_{23}\tau=200 cm^{-1}$,
$|\Om_{12}\tau|=5$, $|\Om_{13}\tau|=20$,$\ga_1\tau=0.02$, $\ga_2\tau=\ga_3\tau=2$
using both Eq.(10) and numerical integration of Eq.(8).
As will be seen later the approximate analytical solution is in excellent agreement with the
numerical evaluation and the ratio of $\Om_{20}(z,t)/\Om_{30}(z,t)$\ at $z=10$\ cm
is equal to a constant (at all t where the two quantities are greater than
$10^{-4}$\ of their peak values) to an accuracy of seven significant figures.
To this accuracy, the constant ratio is  equal to $\Om_{12}/\Om_{13}$.
Also, the peak in the two half-Rabi frequencies occurs at the point predicted
by the group velocity $V_{g1}$.  Note that if one allows $|\Om_{13}|=0$\, then
$V_{g1}$\ reduces to the expression appropriate to the extremely slow wave
propagation experiment [8], as should be the case.

\vskip 10pt
It should be pointed out that the destructive interference predicted above is
equivalent to EIT for both the four-wave mixing photons and the probe laser photons.
To see this clearly, consider the condition for Eq. (1b) to predict $A_2=0$ at
a point in space.  This would be the case if
$A_1=-\Om_{20}/\Om_{21}$ at all times.  With this amplitude for $A_1$\ the coupling
terms destructively interfere.  This is what happens when EIT occurs for the
probe laser, for in this case there is no polarization to lowest order at $\om_p$.
Correspondingly, in order to have EIT at the four-wave mixing frequency we must
have $A_3=0$.  Looking at Eq. (1c), we see that this requires $A_1=-\Om_{30}/\Om_{31}$.
In order for EIT to occur at both frequencies, the two values of $A_1$\ must be
the same.  This yields $\Om_{20}/\Om_{21}=\Om_{30}/\Om_{31}$.  The same relation
holds in Eqs. (10).

\vskip 10pt
We now consider the situation where $|\Om_{12}|\tau=|\Om_{13}|\tau>100$,
$|\de_1|=0$\ and $\ga_1\tau<<1$.  We also assume that $|\de_3|\tau>>1$, but $|\de_2/\Om_{12}|^2, |\de_3/\Om_{12}|^2,
|\de_2/\Om_{13}|^2, |\de_3/\Om_{13}|^2\le 1$. These conditions indicate
that both the Autler-Townes splittings and the detunings are large enough so
that very little absorption occurs.  In this limit three-photon destructive
interference no longer occurs, and we find
\begin{subequations}
\begin{eqnarray}
\Om_{30}(z,t)&=&\frac{\ka_{03}\Om_{31}\Om_{12}}{\ka_{02}|\Om_{13}|^2+\ka_{03}|\Om_{12}|^2}\left(\Om_{20}(0,t-\frac{z}{V_{g1}})-\Om_{20}(0,t-\frac{z}{V_g})e^{iPz}\right), \\
\Om_{20}(z,t)&=&\frac{\ka_{03}|\Om_{12}|^2}{\ka_{02}|\Om_{13}|^2+\ka_{03}|\Om_{12}|^2}\\
&&\quad\times\left(\Om_{20}(0,t-\frac{z}{V_{g1}})+\frac{\ka_{02}|\Om_{13}|^2}{\ka_{03}|\Om_{12}|^2}\Om_{20}(0,t-\frac{z}{V_g})e^{iPz}\right),\nonumber
\end{eqnarray}
\end{subequations}
where we have introduced notations
\begin{eqnarray}
P&=&\frac{\ka_{02}|\Om_{13}|^2+\ka_{03}|\Om_{12}|^2}{|\Om_{12}|^2\de_3+|\Om_{13}|^2\de_2},\nonumber\\
\frac{1}{V_g}&=&\frac{1}{c}+\frac{(1+|\Om_{13}/\Om_{12}|^2)(\ka_{02}|\Om_{13}/\Om_{12}|^2+\ka_{03})}{\left(\de_3+|\Omega_{13}/\Om_{12}|^2\de_2\right)^2}.\nonumber
\end{eqnarray}

\vskip 10pt
In the expression of $V_g$\ we have neglected a term that is of the same order of magnitude as $1/V_{g1}-1/c$, since such a term
is much smaller than $1/V_{g}-1/c$\ in the above equation when the half-Rabi frequencies are much larger than
the detunings. Eq.(11) indicates that there are two contributions to the growth of the FWM field
$\Om_{30}$. The first contribution is due to the probe field that travels
at the group velocity $V_{g1}$, whereas the second term consists of a probe
field that travels at a second group velocity, $V_g$.  If these two parts do not
separate appreciably before a distance $z$ such that $|Pz|=\pi$\ is reached,
the two parts will interfere constructively. In the case where $|\Om_{13}/\Om_{12}|\simeq{1}$,
this requires that $c(\ka_{02}+\ka_{03})<<(\de_2+\de_3)^2$\ and $c/V_{g1}\simeq 1$.
The latter condition puts restrictions on $ka_{12}c\tau^2$\ and $\ka_{32}c\tau^2$,
as compared with $|\Om_{12}\tau|$\ and $|\Om_{13}|\tau$.
\vskip 10pt
The photon flux
conversion efficiency under this condition can be found as
\begin{equation}
\frac{F_m}{F_p}=\frac{\ka_{03}|\Om_{12}|^2\ka_{02}|\Om_{13}|^2}{(\ka_{03}|\Om_{12}|^2+\ka_{02}|\Om_{13}|^2)^2}\left|e^{-(t-z/c)^2/\tau^2}-e^{iPz-Qz-(t-z/V_g)^2/\tau^2}\right|^2,
\end{equation}
where
\begin{equation}
Q=P\frac{|\Om_{12}|^2\ga_3/2+|\Om_{13}|^2\ga_2/2}{|\Om_{12}|^2\de_3+|\Om_{13}|^2\de_2}.\nonumber
\end{equation}
In Eq.(12), $F_m$\ and $F_p$\ are the photon fluxes for the FWM and probe fields,
respectively, and in deriving this expression we have, for the mathematical
simplicity, assumed a Gaussian pulse shape for the probe field so that
$W_{20}(0,\om)=\frac{\Om_{20}(0,0)\tau}{\sqrt{2}}e^{-\om^2\tau^2}$.
>From Eq.(12), we notice that if we choose $|Pz|=\pi$\ and $\ka_{02}|\Om_{13}|^2=\ka_{03}|\Om_{12}|^2$,
then we obtain $F_m\simeq F_p$.  This implies that the conversion efficiency
is close to 100\% whenever the difference between $V_{g1}$\ and $V_g$\ is small
and the absorbtion factor $|Qz|<<1$.  Extensive numerical calculations have verified
that conversion efficiencies of nearly 100\% are indeed produced when
the conditions stated before are well satisfied.

\vskip 10pt
In the case where the three-photon destructive interference does
occur, i.e. in the case when perfect one-photon EIT is established with
$\de_1\tau=\de_2\tau=\de_3\tau=0$, the second term in Eq.(12)
disappears and the integral can be evaluated analytically.  The result is
a nearly 25\% conversion efficiency in spite of the fact that the FWM field
is strongly suppressed and there will be no further production of the field
in the rest of medium.  This indicates that for the scheme studied here, achieving
one-photon EIT is detrimental to the high flux conversion efficiency.  Therefore,
one prefers to detune the two long pulsed lasers from the perspective resonances
in order to avoid the destructive interference.  Another important conclusion
that can be immediately deduced from Eq.(9) is that it shows that maximum atomic
coherence is not the optimum choice for maximum conversion efficiency as
previously suggested [2].  Indeed, there
are two conditions that must be met in order to achieve maximum conversion
efficiency.  The first one is the combination of $\ka_{02}$, $\ka_{03}$,
$|\Om_{12}|$, and $|\Om_{13}|$ that maximize the amplitude of the expression
in Eqs.(10-12).  It has been shown [7] that this is precisely the conventional
phase-matching condition for efficient wave mixing process.  The second
condition that must be met is that the propagation distance must be chosen
properly in order to make the two terms in Eq.(12) interfere constructively. It
is the combination of these two conditions that enables a highly efficient
wave mixing and flux conversion process.
\section{Numerical simulations}
In this section, we investigate the model system numerically.  Our main focus
is to study the validity of Eq.(12) under specified conditions.  This expression
gives a complete description, under the conditions prescribed, on the flux
conversion efficient with the field propagation effect included.  It contains
all the functional dependence of the FWM field on time, propagation distance,
Rabi frequencies and detunings.  We will first compare Eq.(12) with direct numerical
evaluation of the inverse transform in order to show the validity of our approximation
that leads to analytical results.  After having established the validity of the
analytical result, we then investigate the functional dependence of the
conversion efficiency on parameters such as detunings and pumping Rabi frequencies,
which may be useful in experimental verification of the theory.
\vskip 10pt
As mentioned before, the probe field propagation, generation of the FWM field,
and the high-efficiency photon flux conversion described in the previous sections
have all been verified through extensive numerical calculations.  In Figure 2a,
we first show, for a typical set of parameters, the conversion
efficiency predicted by the approximation (Eq.(12)) and by direct numerical
integration of the inverse transform as a function of $(t-z/c)/\tau$. The parameters
are so chosen that the three-photon destructive interference is defeated,
therefore, both terms in Eq.(12) contribute to the overall conversion efficiency.
Furthermore, we have chosen the parameters that maximizes the amplitude of the
expression given in Eq.(11).  The
graph shows an excellent agreement between the approximate solution and the numerical evaluation of the inverse transform, and both methods
predict nearly 100\% conversion efficiency.  In Figure 2b, a different set of propagation
parameters are chosen under the condition where the three-photon destructive
interference is not present.  Again, the plot shows an excellent agreement between
the two methods.  These results indicate that the analytical result Eq.(12) can,
under the conditions given, correctly predict the propagation effect of the
generated field.  Therefore, we will use Eq.(12) to explore the functional dependence
of the conversion efficient to various detunings and power densities.  As has
mentioned before, EIT process is established when both long pulse laser are
tuned on resonance, i.e. $\de_2\tau=\de_3\tau=0$.  The resulting conversion
efficiency can be analytically obtained.  In Figure 3a,
we show the dependence of the efficiency as function of $\de_2\tau$.  As the
detuning of the second laser from the resonance is increased, the efficiency
increases, as expected.  The similar effect due to the detuning $\de_3\tau$ is also shown
in Figure 3b.  These figures indicate that for optimum operation, one should
choose non-vanishing detunings which also satisfy the conditions forth set for
the Rabi frequencies.  Finally, we investigate the dependence of the
conversion efficiency on laser powers.  First, it is obvious that Eq.(9) is
independent of the power of the probe laser.  In Figure 4, we plot the
efficiency as function of $|\Om_{13}/\Om_{12}|^2$. As expected, the maximum
conversion efficiency is achieved only at $|\Om_{13}/\Om_{12}|^2=0.25$,
a necessary condition, for a given $\ka_{02}/\ka_{03}=4$, in order to achieve
phase matching for the generated weave.

\vskip 10pt

\leftline{\bf{V. Conclusion}}
\noindent In conclusion, we have presented an approximate analytical solution to a four-level double-$\Lambda$
scheme where laser induced transparency is expected.  We predict that when
the detunings from the three-photon resonance is small, and therefore the usual
one-photon EIT process is established, the FWM field
propagates without suffering any pulse distortion.
Under these conditions, We also predict
that at a sufficient propagation depth a complete three-photon destructive
interference involving the FWM field and the three laser fields will reduce
the polarization at the four-wave mixing frequency to zero.  We have
shown that even with such a robust three-photon destructive interference that
strongly suppresses further generation of the FWM field beyond
an onset propagation distance, the photon flux conversion efficiency
can still be as high as 25\%.  Our result, however, indicates that one should detune
both pump lasers from the perspective resonances, thereby avoid the usual
one-photon EIT process. The use of a significant detuning will defeat a possible
three-photon destructive interference (that limits the further production of
the internally generated wave and therefore limits conversion efficiency to
25 \%) therefore,  allowing photon flux conversion efficiencies close to 100\%
with even {\it{a very weak probe field}}.  In addition,
our calculation shows that, contrary to common belief,
a maximum atomic coherence between the two lower states is not the optimum
condition for achieving maximum conversion efficiency.  Indeed, with
the current scheme $|\rho_{01}|<<1$.  For high conversion
efficiency, proper atomic parameters should be chosen according to Eqs.(11) and (12).

\section*{List of References}
\begin{itemize}
\item[1.]{S.E. Harris, Phys. Today {\bf{50}}, 36 (1997).}
\item[2.]{A.J. Merriam et al, Phys. Rev. Lett. {\bf{84}}, 5308 (2000),
A.J. Merriam et al, IEEE J. SEL. TOP. QUANT. {\bf{5}}, 1512 (1999),
A.J. Merriam et al, Opt.Lett {\bf{24}}, 625 (1999),
M. Jain et al, Phys. Rev. Lett. {\bf{77}}, 4326 (1996).}
\item[3.]{M.D. Lukin et al, Phys. Rev. A {\bf{60}}, 3225 (1999).  For more
works related to double-$\Lambda$ scheme, see
M.D. Lukin, P.R. Hemmer, and M.O. Scully, Adv. in Atm. Mol. and Opt. Phys., Vol 42, 347 (2000) and reference therein. }
\item[4.]{Early studies on double-$\Lambda$ system coupled with laser excitations
can be found in S.J. Buckle et al., Opta. Acta 33, 1129 (1986).  This study
assumed that four externally generated laser fields, with their phase properties being precisely
controlled, couples all four levels, therefore, no generated field is involved.
It is not surprising that with precise control of the phase of all laser field,
a two-photon cancellation is predicted
by this study.  Since this steady-treatment does not include
the light propagation, therefore could not predict any propagation based
interference effect as demonstrated in the present study.  The same system
was later re-examined by E.A. Korsunsky et al., Phys. Rev. A60, 4996 (1999) with
the extension to include the steady-state Maxwell equation for laser fields.
It also includes discussions on coherent population transfer and induced
transparency.  However, due to its CW nature, the study could not provide any
dynamics of the system.  In addition, the treatment is valid only when all
four laser frequencies are close to each other.  Both studies could not predict,
due to the CW nature of the treatments, a possible three-photon destructive
interference that must involve a phase matched internally generated field.}
\item[5.]{This relation is similar to that obtained in the two references listed
in Ref.[4].  However, it is important to understand the difference between
our results and that given in Ref.[4].  For a four-level system coupled with
four lasers with precise phase control, as described in Ref[4], one can make
such destructive interference work for a wide range of detunings.  In the case
where an internally generated wave contribute to the interference, however, the
robust cancellation effect
can happen only when the phase matching condition for the efficient generation of the
mixing wave is fulfilled.  Further, it is fundamentally different that in the
case where the generated field is excluded, as in Ref. [4], there is no appreciable
effect due to wave propagation, therefore can only be achieved at very low
concentration such as atomic beam.  In the case where the mixing wave participates
in the process, the destructive interference is readily observable
even at elevated concentrations.}
\item[6.]{M.G. Payne, L. Deng, and W.R. Garrett, Phys. Rev. A {\bf{58}},
1361 (1998); L. Deng, M.G. Payne, and W.R. Garrett, Phys. Rev. A {\bf{52}}, 489
(1995); R.C. Hart et al., Phys. Rev. A {\bf{46}}, 4213 (1992); M.G. Payne et al.,
Phys. Rev. A {\bf{44}}, 7684 (1991); J.C. Miller et al., Phys. Rev. Lett. {\bf{45}}, 114 (1980).}
\item[7.]{L. Deng, M.G. Payne, and W.R. Garrett, Phys. Rev. A {\bf{63}}, 3811 (2001);
L. Deng, M.G. Payne, and W.R. Garrett, Phys. Rev. A {\bf{64}}, 31802R (2001).}
\item[8.]{L.V. Hau et al., Nature (London) {\bf{397}}, 594 (1999).}
\end{itemize}
\section*{Figure Captions}

\vskip 10pt
\noindent
Figure 1: Energy level diagram showing a four-level
double-$\Lambda$ scheme with relevant laser
couplings. The theory and predictions presented in the text
are expected to hold with minor modifications for other orderings
of the energies of the excited states.

\vskip 20pt
\noindent
Figure 2a:  A plot of conversion efficiency comparing approximate solution and
direct integration of the inverse transform as a function of $(t-z/c)/\tau$.
Parameters used: $|\Om_{12}|\tau=200$,
$|\Om_{13}|\tau=100$, $\ka_{03}c\tau^2=10$, $\ka_{02}c\tau^2=40$, $\ga_1\tau=\ga_2\tau=\ga_3\tau=0.1$,
$\de_1\tau=0$, $\de_2\tau=20$, $\de_3\tau=20$, and $z=3.93$ cm for constructive interference.
The solid line is the approximate solution whereas the solid
circles are full numerical solutions.  Nearly 100\% conversion efficiency
is predicted by both methods.
\vskip 20pt
\noindent
Figure 2b:  Same plot as in Figure 2a except $\de_1\tau=0$, $\de_2\tau=10$,
$\de_3\tau=10$, and $z=1.96$ cm for constructive interference.
The solid line is the approximate solution whereas the solid
circles are full numerical solutions.
\vskip 20pt
\noindent
Figure 3a:  Flux conversion efficiency vs. a dimensionless quantity $\eta=(t-z/c)/\tau$
for a set of different detuning $\de_2\tau$. Open circle: $\de_2\tau=10$, solid triangle:
$\de_2\tau=20$, open diamond: $\de_2\tau=40$, solid circle: $\de_2\tau=60$.  All
other parameters are the same as in Figure 2a.
\vskip 20pt
\noindent
Figure 3b:  Flux conversion efficiency vs. a dimensionless quantity $\eta=(t-z/c)/\tau$
for a set of different detuning $\de_3\tau$. Open circle: $\de_3\tau=10$, solid triangle:
$\de_3\tau=20$, open diamond: $\de_3\tau=40$, solid circle: $\de_3\tau=60$.  All
other parameters are the same as in Figure 2a.
\vskip 20pt
\noindent
Figure 4:  Flux conversion efficiency vs. a dimensionless quantity $\eta=(t-z/c)/\tau$
for a set of different ratio of Rabi frequencies $|\Om_{13}/\Om_{12}|^2$.
Cross: $|\Om_{13}/\Om_{12}|^2=1$, open circle: $|\Om_{13}/\Om_{12}|^2=0.5$,
solid circle: $|\Om_{13}/\Om_{12}|^2=0.25$, and open triangle: $|\Om_{13}/\Om_{12}|^2=0.1$.
All other parameters are the same as in Figure 2a.
\end{document}